\documentclass[sigconf]{acmart}
\usepackage{multirow}
\AtBeginDocument{%
  }

\setcopyright{acmlicensed}

\copyrightyear{2024}
\acmYear{2024}
\setcopyright{rightsretained}
\acmConference[CIKM '24]{Proceedings of the 33rd ACM International Conference on Information and Knowledge Management}{October 21--25, 2024}{Boise, ID, USA}
\acmBooktitle{Proceedings of the 33rd ACM International Conference on Information and Knowledge Management (CIKM '24), October 21--25, 2024, Boise, ID, USA}
\acmDOI{10.1145/3627673.3679730}
\acmISBN{979-8-4007-0436-9/24/10}

\usepackage{pifont}%
\usepackage[noabbrev,capitalise]{cleveref}
\usepackage{bm}

\newcommand{\paratitle}[1]{\vspace{1.5ex}\noindent\textbf{#1}}

\newcommand{\ignore}[1]{}

\begin{document}

\title{Multi-Behavior Generative Recommendation}

\author{Zihan Liu\textsuperscript{$\dagger$}}
\affiliation{%
  \institution{University of California, San Diego}
  \city{San Diego}
  \state{CA}
  \country{USA}}
\email{zil065@ucsd.edu}
\thanks{$\dagger$ denotes equal contribution.}

\author{Yupeng Hou\textsuperscript{$\dagger$}}
\affiliation{%
  \institution{University of California, San Diego}
  \city{San Diego}
  \state{CA}
  \country{USA}}
\email{yphou@ucsd.edu}

\author{Julian McAuley}
\affiliation{%
  \institution{University of California, San Diego}
  \city{San Diego}
  \state{CA}
  \country{USA}}
\email{jmcauley@ucsd.edu}

\renewcommand{\shortauthors}{Liu et al.}

\begin{abstract}

Multi-behavior sequential recommendation (MBSR) aims to incorporate behavior types of interactions for better recommendations. Existing approaches focus on the next-item prediction objective, neglecting the value of integrating the target behavior type into the learning objective. In this paper, we propose \textbf{MBGen}, a novel \underline{M}ulti-\underline{B}ehavior sequential \underline{Gen}erative recommendation framework. We formulate the MBSR task into a consecutive two-step process: (1) given item sequences, MBGen first predicts the next behavior type to frame the user intention, (2) given item sequences and a target behavior type, MBGen then predicts the next items. To model such a two-step process, we tokenize both behaviors and items into tokens and construct one single token sequence with both behaviors and items placed interleaved. Furthermore, MBGen learns to autoregressively generate the next behavior and item tokens in a unified generative recommendation paradigm, naturally enabling a multi-task capability. Additionally, we exploit the heterogeneous nature of token sequences in the generative recommendation and propose a position-routed sparse architecture to efficiently and effectively scale up models. Extensive experiments on public datasets demonstrate that MBGen significantly outperforms existing MBSR models across multiple tasks. Our code is available at: \href{https://github.com/anananan116/MBGen}{\color{blue}{https://github.com/anananan116/MBGen}}

\end{abstract}



\maketitle

\section{Introduction}\label{sec:intro}
Multi-behavior sequential recommendation (MBSR) has gained attention for its ability to incorporate various behavior types (\emph{e.g.}, click, purchase, add-to-cart) in addition to conventional sequential recommendation~\cite{DMT, DIPN, MBSTR, PBAT, MBHT}.
In the MBSR problem, the model is given an additional sequence of behaviors alongside the item sequence to predict the next interacted item. There are two lines of existing work. 
The first uses a two-stage aggregation paradigm that groups item sequences into subsequences based on the corresponding behavior types, then the subsequences are encoded separately and then aggregated for prediction~\cite{DIPN, DMT}. 
Another paradigm directly models the complete item sequence, using the behavior types as auxiliary sequential input~\cite{MBSTR, PBAT, MBHT}.

Despite their effectiveness, existing models are primarily trained to predict the next items, neglecting the value of integrating \emph{behavior types} into the training objective. Most previous methods can only predict the next items given a specific target behavior~\cite{MBHT, MBGCN, MATH, xia2021graph}. Some methods extend the model's capability to predict the next items under various behavior types~\cite{MBSTR, PBAT}. However, none of these models can jointly predict the next items and behaviors.
The merits of learning to predict the next behavior types are twofold. First, it helps identify the user's intention, which further benefits in predicting the next items. For instance, after users add an item to their cart, they may proceed to checkout, add another complementary item, or continue exploring. Based on these varying intentions, the recommendations should be adjusted accordingly.
Second, behavior-type prediction enables platforms to introduce new functions to improve user experience. Depending on the user's intention, we could redirect the user to the checkout page, display a pop-up window with complementary items, or guide the user back to the home page for further exploration. 

Considering the above discussions, we argue that the MBSR task should be broken down into two consecutive steps. Given the item and behavior sequence, the first step is determining the next behavior. The second step is to predict the next item to interact with, based on both the predicted behavior and the input sequences. However, it's non-trivial to model such a two-step process.
(1) \emph{Two-step dependency.}
An intuitive solution is to follow the multi-task learning paradigm and design an auxiliary behavior-type prediction objective. Although such a multi-task learning approach can simultaneously predict the next items and behavior types, it fails to model the two-step dependency. In the proposed two-step scheme, the item prediction should also take the predicted behavior type as input.
(2) \emph{Varying magnitude of solution spaces.}
Furthermore, the magnitude of the solution space of these two tasks significantly varies. There are usually no more than ten behavior types to predict~\cite{MBSTR,PBAT}. In contrast, the items to predict are usually millions of scale~\cite{zhai2024actions, covington2016deep}, resulting in a much larger solution space. Learning two objectives with varying solution spaces may result in unbalanced learning and lead to sub-optimal performance~\cite{zhang2023advances}.

(3) \emph{Model capabilities.}
In addition, the complex two-step modeling scheme challenges the model's capabilities. Most current MBSR models follow the settings of conventional sequential recommendation and are quite shallow (typically with 2 Transformer layers)~\cite{MBSTR,PBAT,MBHT}. It could be difficult for models with limited capabilities to capture the sequential patterns of both behavior types and items. An effective way to improve model capability is to scale up the model parameters~\cite{kaplan2020scaling}. However, existing literature shows that it is difficult to scale up the model parameters of sequential recommendation models~\cite{zhang2023scaling}. We also observe that existing MBSR models face similar difficulties in scaling up. (See~\cref{fig:scaling_baseline})

\begin{figure*}
\centering
\includegraphics[width=1\textwidth]{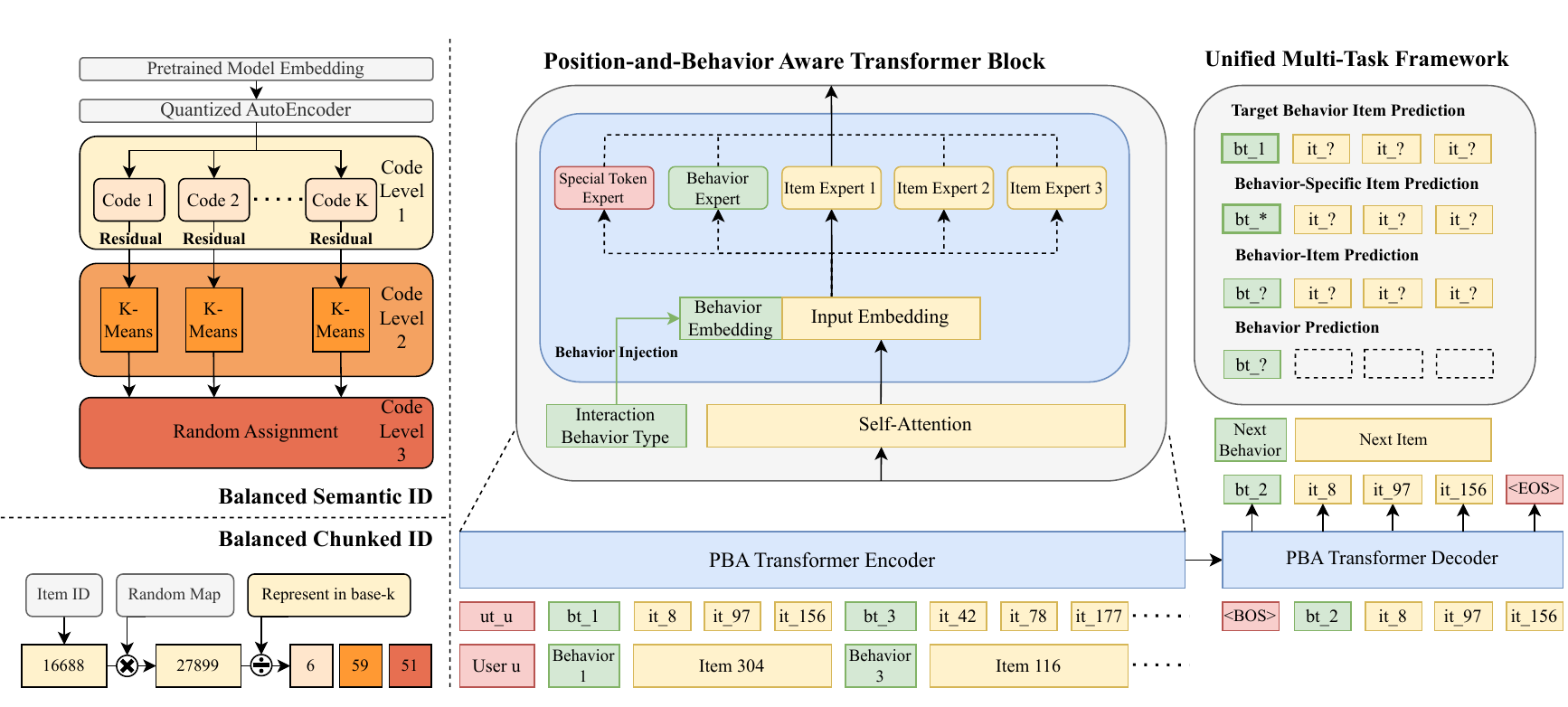}
\vspace{-0.8cm}
\caption{The overall framework of the proposed multi-behavior generative recommendation framework MBGen. We use ``ut'' to denote the user tokens, ``bt'' to denote the behavior tokens, and ``it'' to denote the item tokens.}
\label{fig:model}
\end{figure*}

To this end, we introduce \textbf{MBGen}, the first \underline{M}ulti-\underline{B}ehavior sequential recommendation model under the \underline{Gen}erative recommendation framework. 
To address the imbalanced solution space issue, we tokenize each item into a tuple of ``item tokens'', where item tokens share a much smaller solution space.
We propose modeling the two-step process using a data-centric approach, where behavior tokens and item tokens are flattened and placed interleaved into a single token sequence (\Cref{fig:model}).
This tokenized sequence allows the attention mechanism to capture token-level fine-grained
\texttt{behavior}$\rightarrow$\texttt{item}, \texttt{item}$\rightarrow$\texttt{item}, and \texttt{behavior}$\rightarrow$\texttt{behavior} dependencies.
Furthermore, we can train a sequence-to-sequence generative model with a unified next-token prediction objective.
MBGen naturally learns to predict the next behavior types and items in an auto-regressive manner.
Such a data-driven approach allows us to model the two-step process in a unified framework. Note that the proposed framework inherently has multi-task capability. We can provide any target behavior type as the prompt input to condition the item token generation.
MBGen can also jointly predict the next behavior types and items as usual if the target behavior type is not given. Lastly, the constructed token sequence is heterogeneous by nature. We propose a position-routed sparse network to route each input to different expert networks, which efficiently scales up the model and improves performance.

The main contributions of our work are as follows:\\
\indent $\bullet$ We frame the multi-behavior sequential recommendation task as a two-step process and integrate target behavior types into the learning objective.\\
\indent $\bullet$ We present the first data-centric, token-level generative model for the multi-behavior sequential recommendation tasks with a unified next-token prediction objective. \\
\indent $\bullet$ Extensive experiments on public datasets show that our model significantly outperforms previous MBSR models by 31\% $\sim$ 70\%.

\section{Related Work}
\begin{table*}[t]
\centering
\caption{Comparison of different multi-behavior recommendation models on how they extract behavior-aware interaction patterns and their multi-task capability. Some baseline methods follow a two-stage aggregation paradigm and only model the coarse-grained inter-behavior dependencies at the second aggregation stage on the behavior level. Other baselines model behavior on a behavior-item level, where behavior patterns are injected into each position of the item sequences. Our model captures the behavior patterns at a more fine-grained item token level. In terms of sequential information, early models only consider the sequential patterns in behavior-specific subsequences. Thus, inter-behavior sequential information is ignored in such paradigms. Recent models treat the full item and behavior as separate sequential inputs, while our method combines the behavior and item sequence as a unified heterogeneous sequence.}
\label{tab:related}
\resizebox{\linewidth}{!}{
\begin{tabular}{cccccc}
    \toprule
        \multirow{2}{*}{Model} & \multicolumn{2}{c}{Interaction Pattern} & \multicolumn{3}{c}{Multi-Task Capability} \\
        \cmidrule(lr){2-3} \cmidrule(lr){4-6}
        & Behavior Modeling & Sequential Information & Target Behavior& Behavior-Specific & Behavior-Item \\
        \midrule
        MB-GCN~\cite{MBGCN} & Behavior Level & \ding{55}& \ding{51}& \ding{55}& \ding{55} \\
        MATH~\cite{MATH} & Behavior Level& \ding{55}& \ding{51}& \ding{55}& \ding{55} \\
        DMT~\cite{DMT} & Behavior Level& Single Behavior Subsequences& \ding{51}& \ding{51}& \ding{55}\\
        DIPN~\cite{DIPN} & Behavior Level& Single Behavior Subsequences& \ding{51}& \ding{51}& \ding{55}\\
        MBHT~\cite{MBHT} & Behavior-Item Level& Behavior-Aware Item Sequence& \ding{51}& \ding{55}& \ding{55}\\
        MB-STR~\cite{MBSTR} & Behavior-Item Level& Behavior-Aware Item Sequence& \ding{51}& \ding{51}& \ding{55}\\
        PBAT~\cite{PBAT} &  Personalized User Level& Behavior-Aware Item Sequence& \ding{51}& \ding{51}& \ding{55}\\
        HSTU~\cite{zhai2024actions} & Behavior-Item Level & Behavior-and-Item Sequence& \ding{55}& \ding{55}& \ding{51}\\
        MBGen (ours) & Fine-Grained Token Level & Behavior-and-Item Sequence& \ding{51}& \ding{51}& \ding{51}\\
        \bottomrule
\end{tabular}}
\end{table*}
\paratitle{Sequential Recommendation.}
Sequential recommendation~\cite{hidasi2015session, SASRec} predicts the next item that a user interacts with given the user's interaction history. Early attempts adopt probability models like Markov chains~\cite{rendle2010factorizing}, while modern approaches adopt models like Recurrent Neural Networks (RNN)~\cite{GRU4Rec, hidasi2015session}, Graph Neural Networks (GNN)~\cite{chang2021sequential, wu2019session}, and Convolutional Neural Network (CNN)~\cite{Caser}. More recently, due to its effective attention mechanism in sequential modeling tasks~\cite{dosovitskiy2020image, touvron2023llama}, Transformer~\cite{vaswani2017attention}-based recommendation models~\cite{zhou2020s3, Bert4Rec, SASRec, zhang2019feature, hou2022core} achieves best performance in sequential recommendation tasks. SASRec~\cite{SASRec} uses Transformer decoder as a sequential model, while BERT4Rec~\cite{Bert4Rec} adopts bi-directional attention and masked token prediction as a training objective. However, conventional Transformer-based recommendation models do not incorporate the behavior type of each interaction, which is crucial for understanding the intentions behind user interactions.

\paratitle{Generative Recommendation.}
A new generative recommendation paradigm shows better performance and scalability than the conventional Transformer-based recommender systems on sequential recommendation tasks. In such generative paradigms, each item is tokenized into discrete tokens in an autoregressive manner. These tokens will be further decoded into predicted items. Item tokenization methods are crucial for the performance of generative recommendation methods~\cite{TIGER, petrov2023generative, petrov2024recjpq, jin2023language, liu2024mmgrec}. A common practice is to leverage the text features associated with each item. These text features will then be encoded into
semantically rich item tokens to facilitate the model's decision process. TIGER~\cite{TIGER} generates item tokens with sentence embedding~\cite{devlin2018bert, ni2021sentence} through RQ-VAE~\cite{zeghidour2021soundstream} using descriptive text associated with each item. GPTRec~\cite{petrov2023generative} quantizes item representations from pretrained Truncated Singular Value Decomposition (SVD) models to generate item tokens. \citet{jin2023language} proposed a language model based indexer that could be applied to multiple information retrieval tasks. Such item tokenization techniques also make the generative paradigm more feasible to integrate other modalities, \emph{e.g.}, text, image, and audio~\cite{wang2024enhanced, li2023prompt, zheng2023adapting, wang2024llm, geng2022recommendation}. 

A very recent work, HSTU~\cite{zhai2024actions}, utilizes actions (analogous to behaviors) to unify the retrieval and rank tasks in generative recommender systems. Though similar in the format of the proposed behavior tokens, we highlight two key differences between HSTU and the proposed MBGen: (1) HSTU lacks multi-task capability due to its design that expands behavior tokens after the item tokens. As a result, HSTU cannot predict the next items under a specific target behavior, which is a crucial task in the multi-behavior recommendation scenarios. (2) The behavior plays different roles. HSTU primarily uses actions as a preference indicator for ranking tasks and adopts a target-aware retrieval and ranking paradigm. In contrast, we model the MBSR task as a consecutive two-step process (as discussed in~\Cref{sec:intro}). The prediction of the next behaviors helps MBGen to reveal user intentions for better recommendations.

\paratitle{Multi-Behavior Recommendation.}
Multi-behavior recommendation models incorporate behavior types of history user-item interactions to predict the items that the user will interact with. Early methods mainly exploit MF-based models for multi-behavior recommendations~\cite{krohn2012multi, zhao2015improving, singh2008relational}. Later, some deep learning methods~\cite{chen2020efficient, gao2019neural}, especially GNN-based methods~\cite{chen2021graph, xia2021graph, zhang2020multiplex, MBGCN, yang2022dpgnn}, are also proved to be effective. However, the above-mentioned methods do not include temporal sequential patterns in their model, which is vital for next-item prediction.

Recent works have started to study the multi-behavior sequential recommendation (MBSR) problem. Early MBSR models~\cite{DMT, DIPN, chen2021curriculum} usually take sequential input of each behavior as separate subsequences and aggregate the representation encoded from each subsequence to predict the next interaction. However, this aggregation method ignores the sequential pattern between different behaviors. Thus, other approaches aim to directly integrate the sequence of behaviors into the item sequence~\cite{MBSTR, PBAT, MBHT}. MB-STR~\cite{MBSTR} routes each input item to different sets of feed-forward and linear transformation layers inside the Transformer model according to the corresponding behavior type. PBAT~\cite{PBAT} considers the user's personalized interaction pattern of behaviors. MBHT~\cite{MBHT} uses hypergraph learning to uncover item-wise multi-behavior correlations. While effective, as shown in \Cref{tab:related}, previous works do not treat next behavior type prediction as one of their training objectives and thus fail to capture users' intention in the MBSR problem. In this work, we aim to predict both the next behavior and the next item in a unified generative recommendation framework.

\section{Methods}
    In this section, we present MBGen, a framework designed to seamlessly integrate various behavior-related recommendation tasks into a single framework. In~\Cref{sec:formulation}, we first formally define the three multi-behavior recommendation tasks. Then, in~\Cref{sec:item_id}, we introduce our tokenizer that maps each raw item ID to multiple item tokens to achieve token-level fine-grained pattern modeling. We then elaborate on our model's capability on multiple multi-behavior recommendation tasks in~\Cref{sec:multi-task}. We finally present the proposed position-and-behavior-aware Transformer block in~\Cref{sec:PBA} to better model the heterogeneous token sequences. The overall framework of MBGen is illustrated in~\Cref{fig:model}.
\subsection{Problem Formulation}\label{sec:formulation}
    Multi-behavior sequential recommendation models the next item to interact with, given the historical interaction sequence (in other words, item sequence) and the corresponding behavior types. Formally, we denote $\mathcal{U}=\{u_1,u_2,\ldots,u_{|\mathcal{U}|}\}$ as the set of users and $\mathcal{V}=\{v_1,v_2,\ldots,v_{|\mathcal{V}|}\}$ as the set of items. At the same time, we define a set of behaviors that are associated with each interaction: $\mathcal{B}=\{b_1,b_2,\ldots,b_{|\mathcal{B}|}\}$, among which the most important behavior is marked to be the target behavior, \emph{e.g.}, ``purchase'' is the typical target behavior in e-commerce platforms in comparison to other behaviors like ``view'', ``add-to-cart'', and ``add-to-favorite''. For each user $u$, we define the behavior-aware interaction history sequence to be $S_u = \left[\left<v_1^u, b_1^u\right>, \ldots \left<v_{n-1}^u, b_{n-1}^u\right>\right]$. The user's last interaction $\left<v_{n}^u, b_{n}^u\right>$ is the prediction target. As discussed in~\Cref{sec:intro}, we propose to formulate the MBSR problem as a two-step process. First, the interaction history $S_u$ determines the behavior type of the next interaction $b_{n}^u$, which helps frame the user intention. Then, $S_u$ and $b_{n}^u$ jointly determine the next interacted item $v_{n}^u$. Thus, we can model the probability distribution of the next behaviors and items as follows:
    \[
    \operatorname{P}(b_n, v_n | S_u) = \operatorname{P}(b_n | S_u) \cdot \operatorname{P}(v_n | S_u, b_n).
    \]

    Then, we describe the three behavior-related tasks in the MBSR problem, \emph{i.e.}, target behavior item prediction, behavior-specific item prediction, and behavior-item prediction. In the target behavior item prediction task, we model $\operatorname{P}(v_n | S_u, b_n)$ given $b_n=b_1$, where $b_1$ is the target behavior for simplicity. Similarly, in the behavior-specific task, the model predict $\operatorname{P}(v_n | S_u, b_n)$ given arbitrary behavior $b_{n} \in \mathcal{B}$.
    In contrast, the behavior-item prediction task gives no information about the next behavior. The model predicts both $\operatorname{P}(b_n | S_u)$ and $\operatorname{P}(v_n | S_u, b_n)$ given $S_u$.

\subsection{Learning Balanced Behavior\&Item IDs}\label{sec:item_id}
    In the presented generative recommendation framework, to balance the solution space of the next behaviors (usually less than ten) and items (millions), we propose to tokenize each item into a tuple of tokens.
    Existing works~\cite{TIGER, liu2024mmgrec} use RQ-VAE~\cite{zeghidour2021soundstream} as the item tokenizer to encode dense item features into tokens. However, we observe that this practice is suboptimal due to the extremely unbalanced code distribution (\Cref{tab:code_distribution}), which necessitates a larger pool of tokens and an additional digit to prevent collisions. In this section, we propose to tokenize items into distribution-balanced tokens. More balanced token distribution leads to smaller token pools and reduced solution spaces, resulting in better performance~\cite{petrov2024recjpq, hou2023learning}.
    We refer to the item tokens that come from associate item features as Balanced Semantic ID (SID). 
    We also propose a balanced item tokenization method that takes only item IDs as inputs, referred to as the Balanced Chunked ID (CID).

    \paratitle{Balanced Semantic ID.}
    For the widely-used RQ-VAE tokenization~\cite{TIGER}, we observe that token imbalances stem from the high correlations between residual vectors and the quantized first-digit token.
    The second and subsequent digits of tokens derived from the residuals should ideally be uniformly distributed, regardless of the first digit of the token. 
    However, existing methods use a global codebook to quantize residual vectors, regardless of which first-digit token the item is assigned.
    Thus, residuals with the same first-digit token are likely to be quantized to have the same second-digit (also subsequent) token.
    With this in mind, we propose individually quantizing the residuals of each first-digit token rather than maintaining a single codebook for all residuals.
    
    \emph{(1) Code level 1.} When deriving the first digit, we fit a quantized auto-encoder with the item features as input.
    This process can also be seen as an RQ-VAE \cite{zeghidour2021soundstream} with only one quantization layer.
    We first feed the item feature $\bm{x}$ into an encoder $E(\cdot)$ (a simple feed-forward network in our case) to learn the latent representation $\bm{z} \coloneq E(\bm{x}) \in \mathbb{R}^d$, where $d$ is the dimension size. 
    Meanwhile, we maintain a codebook $\mathcal{C} \coloneq \bm{E} \in \mathbb{R}^{K\times d}$, where 
    $K$ is the codebook size.
    We then find the closest codebook entry and use the corresponding entry index $r$ as the first-digit token, \emph{i.e.}, 
    $r = \operatorname{argmin}_j \| \bm{z} - \bm{e}_j\|$.
    We then use another neural network as a decoder to 
    reconstruct the input item feature: $\hat{\bm{x}} \coloneq D(\bm{e}_r)$.
    This network is trained with two losses: (1) the reconstruction loss $\mathcal{L}_{rec} = \|\hat{\bm{x}} - \bm{x}\|^2$ and, (2) the quantization loss $\mathcal{L}_{quan} = \| \bm{z} - \operatorname{sg}\left[ \bm{e}_r \right] \|^2$, where $\operatorname{sg}[\cdot]$ denotes the stop gradient operation. The final loss is given by $\mathcal{L} = \mathcal{L}_{rec} + \beta \mathcal{L}_{quan}$, where $\beta$ is the hyperparameter to balance the two losses. The codebook is maintained by running average as proposed in Razavi \emph{et al.}~\cite{razavi2019generating}. The residual $\bm{z} - \bm{e}_r$ will be used for deriving the second-digit token.
    
    \emph{(1) Code level 2 and 3.} When deriving the second digit, 
    for every item that shares the first digit, we fit an individual k-means model with $K$ as the number of clusters.
    Then the index of each cluster is used as the second-digit token. 
    By fitting individual k-means models for residuals that share the same first-digit token, we can generate more balanced second-digit tokens. Such balanced distribution allows us to tokenize the same number of items with fewer codebook layers and smaller codebook sizes. In this way, the proposed balanced semantic ID (SID) tokenization method reduces the model's decision space and computational consumption due to the shorter input sequences~\cite{petrov2024recjpq}. Eventually, we randomly assign the third-dight token for every item with the same first and second digits to resolve the collision issue and ensure every SID uniquely identifies an item.

    \paratitle{Balanced Chunked ID.}
    We also propose a tokenizer that does not require any associated item features. For a given item ID, we first remap it into one random integer $i$. To reduce the space size $|\mathcal{V}|$ of these integers, we represent the integer in base \(k\). This process ensures that each item is represented by a tuple of digits, where each digit serves as a token. Although this approach lacks semantic information from item features, it still generates effective item tokens by ensuring a balanced token distribution.
    
    \paratitle{<behavior, item> Tokenizer.}
    We use a separate behavior token to encode each type of behavior and add the behavior token before the item tokens to encode behavior information in the multi-behavior interaction sequence. Overall, we will tokenize each item $v\in \mathcal{V}$ into a tuple of tokens $C^v \coloneq \left[ c^v_1, c^v_2, \ldots, c_m^v\right]$ that uniquely identifies each item $v$. Combining the behavior token $b$, each interaction can be represented as the tuple $\left[b, c^{v}_1, c^{v}_2, c_3^{v}\right]$. In this way, the above-mentioned MBSR problem has been formalized as a \emph{next-token prediction} task. Given the user interaction history $S_{u}' = \left[b^{v_1}, c^{v_1}_1, c^{v_1}_2, c_3^{v_1}, \ldots, b^{v_{n-1}}, c^{v_{n-1}}_1, c^{v_{n-1}}_2, c_3^{v_{n-1}}\right]$, the task is to predict the next interaction $\left[b^{v_n}, c^{v_n}_1, c^{v_n}_2, c_3^{v_n}\right]$ token-by-token.
    
\subsection{Generative Multi-Task Multi-Behavior Modeling}\label{sec:multi-task}
    \paratitle{Sequence-to-Sequence Generative Framework.}
    To solve the above-mentioned two-step MBSR problem, we first tokenize the original item and behavior sequence into one single token sequence using the item and behavior tokenizer proposed in~\Cref{sec:item_id}. Then, the token sequences are fed into an encoder-decoder Transformer~\cite{vaswani2017attention} module to generate the next tokens autoregressively.
    When the decoder is trained to give the next token prediction, it naturally learns the two-step prediction objective in the MBSR problem, \emph{i.e.}, first predicts the next behavior to frame user intention, then predicts the next item given the next behavior type. Additionally, to capture the personalized sequential and behavior interaction pattern, we follow \citet{TIGER} and map the raw user IDs to $U$ tokens using the hashing trick \cite{weinberger2009feature}. The mapped user ID is prepended to the input token sequence to help capture the personalized interaction features on the encoder side.
    
    \paratitle{Multi-Task Capability.}
    Learning from the proposed token sequences naturally gives MBGen multi-tasking capabilities.
    As mentioned, we formulate the MBSR problem into three tasks: target behavior item prediction, behavior-specific item prediction, and behavior-item prediction. For both target behavior item prediction and behavior-specific item prediction tasks, we prompt the sequence model with $\left[ \left<\texttt{BOS}\right>, b\right]$, where $b$ is the target behavior token in target behavior item prediction task, and can also be an arbitrary behavior token in behavior-specific item prediction task. In this way, we effectively condition the generation of the following item tokens using the provided behavior token. For the behavior-item prediction task, the model is only provided with the $\left<\texttt{BOS}\right>$ token and is responsible for generating both the behavior token and the item tokens. To the best of our knowledge, MBGen is the first model that can make predictions on the next behavior and the behavior-specific next-item prediction simultaneously. 
\subsection{The Position-and-Behavior-Aware Transformer}\label{sec:PBA}

    \paratitle{Position Routed Mixture of Experts.}
    Note that the input token sequences described in~\Cref{sec:item_id} are heterogeneous by nature.
    For example, the first token in each behavior-item tuple always encodes the behavior information.
    Inspired by the Mixture of Expert (MoE) architecture~\cite{fedus2022switch, jiang2024mixtral}, we propose to set up multiple expert networks to model the sequence heterogeneity. The input of each digit from a behavior-item tuple $\left[b, c_1, c_2, c_3\right]$ will be routed to different expert networks.
    Existing MoE models typically use a trainable gating network to route the input to each expert network. It's known that such a gating technique usually suffers from problems like representation collapse~\cite{chi2022representation}.
    Thus, we propose to route the inputs purely by the relative position of each token.
    In detail, we duplicate the Feed Forward Network (FFN) in the Transformer blocks.
    The behavior token and three item tokens are routed to four different FFN experts, respectively.  All other tokens (\emph{e.g.}, $\left<\texttt{}{EOS}\right>$, $\left<\texttt{BOS}\right>$, user token) are routed to another FFN expert. Another merit of the proposed position routed module is that it efficiently scales up the model parameters without additional computational budget.
    
    \paratitle{Behavior Injected Feed Forward Layer.}
    To further enhance the model's capability of understanding the fine-grained behavior-item interaction, we design an additional behavior embedding table in the FFN layers. Each FFN layer accepts an additional input of the current behavior type. The behavior types for the item tokens are decided by the associated behavior token, and the behavior types for the behavior tokens and the special tokens are set to be the padding token. The corresponding behavior embedding is then concatenated with the input of the FFN layer to facilitate a better mix of item and behavior information.

\section{Experiment}
\subsection{Experiment Setup}
    \begin{table}[!t] %
        \caption{Statistics of the preprocessed datasets. ``Avg.~$n$'' denotes the average length of item sequences. ``Avg.~$w$'' denotes the average number of words in the item text.}
        \label{tab:dataset}
        \resizebox{\columnwidth}{!}{
        \begin{tabular}{l *{5}{r}}
            \toprule
            \textbf{Datasets} & \textbf{\#Users} & \textbf{\#Items} & \textbf{\#Inters.} & \textbf{Avg. $n$} & \textbf{Avg. $w$}\\
            \midrule
            \textbf{Retail}  &  147,894 &  97,842 &  7,651,493 & 78.20 & 51.74 \\
            \textbf{IJCAI}      & 423,119 &  351,221 & 32,685,371 & 93.06 & 77.25 \\
            \bottomrule
        \end{tabular}
        }
    \end{table}
    \subsubsection{\textbf{Datasets}}
    We evaluate our proposed MBGen model with the following two public datasets\cite{xia2021graph}:

    \textbf{Retail.} The retail dataset is collected from Taobao, one of the world's largest e-commerce platforms. There are four types of behaviors: purchase, add-to-cart, add-to-favorite, and click. The purchase behavior is seen as the target behavior in this dataset. Some baselines also refer to this dataset as ``Taobao''/``Tmall''.

    \textbf{IJCAI.} This dataset was used for the IJCAI competition and contains real-world data collected from e-commerce platforms. Similar to the retail dataset, there are four types of behaviors: purchase, add-to-cart, add-to-favorite, and click. The purchase behavior is seen as the target behavior in this dataset.

    We filter all the datasets to exclude the items that appear less than 5 times in the training sequence. The statistics of the filtered datasets are shown in~\Cref{tab:dataset}. We do not use the Yelp dataset used as in~\cite{MBSTR, PBAT}. In the Yelp dataset, the user rating field (on a scale of 1 (worst) to 5 (best)) is split into three types of behaviors: dislike ($\text{rating} \leq 2$), neutral ($2 < \text{rating} < 4$), and like ($\text{rating} \geq 4$). However, this doesn't align with the real-life distribution of behaviors.

\begin{table*}[t]
    \small
    \centering
    \caption{Performance comparison of different models on the target behavior prediction task. The best performance is denoted in bold font. We use the underlined font to denote the best performance in baseline models if one variant of our proposed MBGen achieves the best performance. ``*'' indicates that our method statistically significantly (\emph{i.e.}, p-value<0.05 in the paired t-test) outperform the best baseline.}
    \label{tab:main_exp}
    \begin{tabular}{p{2.3cm}ccccccccc}
        \toprule
            \multirow{2}{2.3cm}{\centering Model Type} & \multirow{2}{*}{Model} & \multicolumn{4}{c}{Retail} & \multicolumn{4}{c}{IJCAI} \\
        \cmidrule(lr){3-6} \cmidrule(lr){7-10}
            & & HR@5 & NDCG@5 & HR@10 & NDCG@10 & HR@5 & NDCG@5 & HR@10 & NDCG@10  \\
        \midrule
            \multirow{2}{2.3cm}{\centering Multi-Behavior Recommendation}
            & MB-GCN & 0.0485 & 0.0316 & 0.0732 & 0.0396 & N/A & N/A & N/A & N/A\\
            & MB-GMN & 0.0012 & 0.0008 & 0.0022 & 0.0011 & N/A & N/A & N/A & N/A\\
        \midrule
            \multirow{4}{2.3cm}{\centering Sequential Recommendation}
            & GRU4Rec & 0.2869 & 0.206 & 0.3554 & 0.2282 & 0.2117 & 0.1539 & 0.272 & 0.1734\\ 
            & $\text{BERT4Rec}_\textbf{M}$ & 0.0807 & 0.0670 & 0.0934 & 0.0711 & 0.0617 & 0.0457 & 0.0799 & 0.0516\\ 
            & $\text{SASRec}_\textbf{M}$ & \underline{0.3041} &  \underline{0.2243} &  \underline{0.3678} &  \underline{0.2449} & 0.2294 & 0.1785 & 0.2818 & 0.1954\\ 
            & TIGER & 0.2887 & 0.2049 & 0.3523 & 0.2263 & \underline{0.3728} & \underline{0.2595} & \underline{0.4310} & \underline{0.2788}\\ 
        \midrule
            \multirow{5}{2.3cm}{\centering Multi-Behavior Sequential Recommendation}
            & $\text{BERT4Rec}_\textbf{B}$ & 0.005  & 0.0032 & 0.0076 & 0.0040 & 0.0298 & 0.0207 & 0.0417 & 0.0245\\ 
            & $\text{SASRec}_\textbf{B}$ & 0.1369 & 0.0945 & 0.1788 & 0.1081 & 0.1295 & 0.0950 & 0.1662 & 0.1069\\ 
            & PBAT & 0.1112 & 0.075 & 0.1545 & 0.089 & 0.1497 & 0.1048 & 0.2011 & 0.1214\\ 
            & MBHT & 0.2984 & 0.2183 & 0.3608 & 0.2385 & 0.1652 & 0.1249 & 0.2069 & 0.1384\\ 
                & MB-STR & 0.2814 & 0.2066 & 0.3500 & 0.2289 & 0.1874 & 0.1386 & 0.236 & 0.1544\\ 
        \midrule
        \midrule
            \multirow{3}{2.3cm}{\centering Multi-Behavior Generative Recommendation}  
            & \textbf{MBGen (SID)} & 0.4859$^*$ & 0.3800$^*$ & \textbf{0.5329}$^*$ & \textbf{0.3954}$^*$ & \textbf{0.4895}$^*$ & \textbf{0.3498}$^*$ & \textbf{0.6285}$^*$ & \textbf{0.3949}$^*$\\ 
            & \textbf{MBGen (CID)} & \textbf{0.4886}$^*$ & \textbf{0.3815}$^*$ & 0.5245$^*$ & 0.3934$^*$ & 0.4866$^*$ & 0.3454$^*$ & 0.6281$^*$ & 0.3913$^*$\\ 
            & \# Improve & \textbf{+60.67\%} & \textbf{+70.08\%} & \textbf{+44.89\%} & \textbf{+61.45\%} & \textbf{+31.30\%} & \textbf{+34.45\%} & \textbf{+45.82\%} & \textbf{+41.64}\%\\ 
        \bottomrule
        \end{tabular}
\end{table*}
    
    \subsubsection{\textbf{Baselines}} We compare the proposed approach with the following baseline methods:
    
    \noindent \textbf{(1) Sequential Recommendation Models.}\
    For all sequential recommendation models, we treat the multi-behavior item sequence as a regular single-behavior item sequence to enhance the sequential baseline models.\\
        \indent $\bullet$ \textbf{GRU4Rec}~\cite{GRU4Rec} uses a Gated Recurring Unit (GRU) to enhance long-term memory in recommendation tasks.\\
        \indent $\bullet$  $\textbf{SASRec}_{\textbf{M}}$~\cite{SASRec} utilizes a self-attentive model to capture item correlations. We use the \textbf{M} subscript to denote the usage of multi-behavior sequence, aligning with the notation in previous works~\cite{MBSTR}.\\
        \indent $\bullet$  $\textbf{BERT4Rec}_{\textbf{M}}$~\cite{Bert4Rec} adopts a bi-directional self-attentive model with a cloze objective for sequence modeling. We use the \textbf{M} subscript to denote the usage of multi-behavior sequence, aligning with the notation in previous works~\cite{MBSTR}.\\
        \indent $\bullet$  \textbf{TIGER}~\cite{TIGER} proposes a novel learned Semantic ID representation for each item and achieved State of the Art (SOTA) performance in sequential recommendation tasks.

    \noindent \textbf{(2) Multi-behavior Recommendation Models.}\\
        \indent $\bullet$ \textbf{MB-GCN}~\cite{MBGCN} leverages a unified graph structure to represent multiple types of user-item interactions.\\
        \indent $\bullet$ \textbf{MB-GMN}~\cite{xia2021graph} incorporates a meta-learning paradigm to capture interaction diversity and behavior heterogeneity.

    \noindent \textbf{(3) Multi-behavior Sequential Recommendation Models.}\\
        \indent $\bullet$ \textbf{HSTU}~\cite{zhai2024actions} uses actions to unify retrieval and ranking in the same model. In our reproduction, we include behaviors as ``actions''. In detail, we place the item and behavior tokens interleaved as <item, behavior> for each interaction.\\
        \indent $\bullet$ \textbf{MBHT}~\cite{MBHT} utilizes hypergraph learning to uncover item-wise multi-behavior correlations.\\
        \indent $\bullet$ \textbf{PBAT}~\cite{PBAT} takes the user’s personalized behavior interaction pattern into account using learned user-associated features.\\
        \indent $\bullet$ \textbf{MB-STR}~\cite{MBSTR} adopts a sparse Mixture of Experts (MoE) architecture and sequential pattern generator to handle the semantic gap between different behaviors.\\
        \indent $\bullet$ $\textbf{SASRec}_{\textbf{B}}$~\cite{SASRec} and $\textbf{BERT4Rec}_{\textbf{B}}$~\cite{Bert4Rec} are enhanced from the original sequential recommendation methods by seeing each item with a different behavior type as a new item, thus building a new set of items of size $\vert\mathcal{V}\vert \times \vert\mathcal{B}\vert$. The enhanced models gain the ability to perform both behavior-specific and behavior-item prediction tasks. We denote the enhanced version of SASRec as $\textbf{SASRec}_{\textbf{B}}$, and the enhanced version of BERT4Rec as $\textbf{BERT4Rec}_{\textbf{B}}$.

    \subsubsection{\textbf{Evaluation Settings}}
    Following previous works~\cite{MBSTR, PBAT, MBHT}, we adopt two widely used ranking metrics, Recall@K and NDCG@K, where K $\in\{5,10\}$. For dataset splitting, we apply the leave-one-out strategy, \emph{i.e.} the latest interacted item as test data, the item before the last one as validation data. Different from previous works~\cite{MBSTR, PBAT, MBHT}, the ground-truth item of each sequence is ranked among \emph{all the other items} instead of some sampled negative items for reliable evaluation~\cite{krichene2020sampled}.
    We conduct a beam search with 50 beams to rank the top 10 candidates in all items to evaluate the Recall@K and NDCG@K metrics for the proposed MBGen model. All item sequences are truncated to a length of $50$ for training and evaluation of our proposed model and the baselines.

    \subsubsection{\textbf{Implementation Details}}\
        We reproduce most of the baselines (apart from MB-GMN) in the RecBole~\cite{zhao2021recbole, zhao2022recbole} framework and use the same truncated item sequence to train all the models. 
        
        \textbf{Item Tokenizer.} We use a vocabulary size of $(K \times m)$ where $m = 3$ for all datasets. $K = 96$ for the IJCAI dataset, $K = 64$ for the retail dataset. Unlike previous works~\cite{TIGER, jin2023language, liu2024mmgrec} that require associated item features as additional input, we use the embedding table of a pretrained sequential recommendation model (MB-STR~\cite{MBSTR} in our case) as the item feature and input of semantic ID generation. In the first quantization phase of balanced SID, the Quantized Auto-Encoder has hidden layers of size $\left[2048, 1024, 512, 256\right]$ in its encoder and decoder with the ReLU activation function. The latent dimension is set to be 32. The codebook size is the same as $K$ for each dataset. We use a batch size of $2048$, a learning rate of $0.001$, and a $\beta$ of $0.25$ to train the model using the Adagrad optimizer until full convergence. We choose the same $k$ for the two datasets in the generation of balanced CID. 
    
        \textbf{Sequence-to-Sequence Model.} We implement our sequence-to-sequence prediction model based on the Switch Transformers~\cite{fedus2022switch} architecture. We use a model dimension of 256, an inner dimension of 512 with ReLU activation function, and 6 heads of dimension 64 in the self-attention layer. The model has 4 regular encoder layers and 4 sparse decoder layers, and we add behavior injection in the first two encoder and decoder layers. Each sparse layer has 5 of experts in the sparse FFN layer. We set the batch size to $512$, learning rate to $0.001$, and trained our model with the AdamW optimizer for 350000 steps on both datasets. The learning rate is tuned in $\{0.0008, 0.001, 0.0012\}$ and the weight decay is tuned in $\{0.005, 0.001\}$. The models with the best NDCG@10 performance on the validation set are selected to be tested on the test set.

\subsection{Overall Performance}
        
    \subsubsection{\textbf{Target Behavior Item Prediction}}
    We compare the two variants (using SID and CID proposed in~\Cref{sec:item_id}, respectively) of our proposed MBGen with the baseline methods on the target behavior item prediction task and report the results in \Cref{tab:main_exp}. We only use sequences that end with an interaction with target behavior for target behavior evaluation.

    For baseline methods, we observe that all baselines that incorporate sequential information significantly outperform non-sequential baseline models (\emph{i.e.}, MB-GCN, and MB-GMN), verifying that sequential patterns are a crucial part of the MBSR problem. Furthermore, we cannot finish training and evaluating the MB-GCN and MB-GMN models in a reasonable time on the IJCAI dataset since it's not designed for retrieval tasks on such a large-scale dataset. Different from the results reported in previous MBSR works, we observe that some sequential recommendation models (\emph{e.g.}, SASRec, and MB-STR) outperform all the MBSR models. This could be caused by the following reasons: (1) Some previous works~\cite{MBHT, PBAT} train the sequential recommendation baseline models using only the target behavior sequence, which undermines the performance of the sequential recommendation baselines. (2) Previous works~\cite{MBSTR} use the SASRec and GRU4Rec models that was trained with the original binary entropy loss instead of the cross entropy loss, which is known to provide a significant improvement over the original SASRec model~\cite{zhai2023revisiting}. (3) In our evaluation setting, the ground-truth item of each sequence is ranked among all the other items instead of some sampled negative items as in~\cite{MBSTR, MBHT, PBAT}. We also notice that the behavior-enhanced SASRec$_\textbf{B}$ and BERT4Rec$_\textbf{B}$ models perform worse than the original SASRec and Bert4Rec modes. This could be due to the increased sparsity of the data after we create additional items for each behavior type. Additionally, we observe that the generative recommendation method (\emph{i.e.}, TIGER) significantly outperforms conventional sequential recommendation models (\emph{e.g.}, SASRec and GRU4Rec) when retrieving from a large number of candidates in the IJCAI dataset. For the proposed MBGen model, both of its variants significantly outperform all baselines in both datasets by 30\% to 70\%. 
    \subsubsection{\textbf{Behavior-Specific Item Prediction}}
    We compare the proposed MBGen model with the four baselines with the behavior-specific item prediction capability and present the results in~\Cref{tab:behavior_specific_exp}. In this task, the model is given the behavior type of the last interaction that the user performs and is required to predict the interacted items. We note that all models perform worse on behavior-specific item prediction task than target behavior item prediction task. 
    This follows our intuition the target behavior (\emph{e.g.}, purchase) usually indicates a stronger preference than other behaviors (\emph{e.g.}, click). Thus, it could be easier to predict interactions with target behavior than arbitrary behavior. The proposed MBGen model also performs best in the behavior-specific item prediction task among all baselines.
    
    \subsubsection{\textbf{Behavior-Item Prediction}}
    We evaluate the performance of our proposed MBGen and other baselines on the behavior-item prediction task that requires the model to make predictions on both the behavior type and the item of the user's next interaction. We count a prediction as correct if the prediction of both behavior and item matches the ground-truth behavior and item. As shown in~\Cref{tab:behavior_item_exp}, despite its lack of capability on the other two tasks, HSTU performs best among all baseline models. Our proposed MBGen model also outperforms all baseline methods in the behavior-item prediction task. We also note only a 10\%-20\% drop in performance compared to the behavior-specific prediction task, showing our model's solid ability in the next-behavior prediction task in the MBSR problem.
\begin{table}[t]
    \centering
    \caption{Performance comparison on the behavior-specific item prediction task.}
    \label{tab:behavior_specific_exp}
    \resizebox{\linewidth}{!}{
    \begin{tabular}{ccccc}
        \toprule
        \multirow{2}{*}{Model} & \multicolumn{2}{c}{Retail} & \multicolumn{2}{c}{IJCAI} \\
        \cmidrule(lr){2-3} \cmidrule(lr){4-5}
            & NDCG@5 & NDCG@10 & NDCG@5 & NDCG@10 \\
        \midrule
            $\text{BERT4Rec}_\textbf{B}$ & 0.0021 & 0.0013 & 0.0037 & 0.0018 \\ 
            $\text{SASRec}_\textbf{B}$ & 0.0468 & 0.0580 & 0.0822 & 0.0946 \\ 
            PBAT & 0.0346 & 0.0430 & 0.0799 & 0.0944 \\ 
            MB-STR & \underline{0.0606} & \underline{0.0715} & \underline{0.1109} & \underline{0.1257}\\ 
        \midrule
            \textbf{MBGen (SID)} & \textbf{0.2053} & \textbf{0.2131} & \textbf{0.3023} & \textbf{0.3317}\\
            \textbf{MBGen (CID)} & 0.2030 & 0.2076 & 0.2972 & 0.3261\\ 
            \bottomrule
        \end{tabular}
        }
    \end{table}

    \begin{table}[t]
    \centering
    \caption{Performance comparison on the behavior-item prediction task.}
    \label{tab:behavior_item_exp}
    \resizebox{\linewidth}{!}{
    \begin{tabular}{ccccc}
        \toprule
        \multirow{2}{*}{Model} & \multicolumn{2}{c}{Retail} & \multicolumn{2}{c}{IJCAI} \\
        \cmidrule(lr){2-3} \cmidrule(lr){4-5}
            & NDCG@5 & NDCG@10 & NDCG@5 & NDCG@10 \\
        \midrule
            $\text{BERT4Rec}_\textbf{B}$ & 0.0013 & 0.0007 & 0.0024 & 0.0010\\ 
            $\text{SASRec}_\textbf{B}$ & 0.0299 & 0.0395 & 0.0287 & 0.0375\\ 
            HSTU & \underline{0.1579} & \underline{0.1616} & \underline{0.1607} & \underline{0.1654}\\ 
        \midrule
            \textbf{MBGen (SID)} & 0.1712 & \textbf{0.1803} & 0.2155 & 0.2301\\
            \textbf{MBGen (CID)} & \textbf{0.1720} & 0.1800 & \textbf{0.2443} & \textbf{0.2730}\\ 
            \bottomrule
        \end{tabular}
        }
    \end{table}
\begin{table*}[t]
        \centering
        \caption{Ablation analysis on three MBSR tasks. The best performance is denoted in bold font, and underlined font is used to denote the best performance in ablation models if one variant of our proposed MBGen achieves the best performance. We use \textit{PR} to denote the position routed sparse FFN layer, \textit{BI} to denote the behavior injection module, and \textit{IT} to denote the item tokenizer.}
        \vspace{-0.1in}
        \label{tab:ablation_exp}
        \resizebox{\linewidth}{!}{
        \begin{tabular}{ccccccccccccc}
            \toprule
            \multirow{4}{*}{Model} & \multicolumn{6}{c}{Retail} & \multicolumn{6}{c}{IJCAI} \\
            \cmidrule(lr){2-7} \cmidrule(lr){8-13}
            & \multicolumn{2}{c}{Target Behavior} & \multicolumn{2}{c}{Behavior Specific} & \multicolumn{2}{c}{Behavior-Item} & \multicolumn{2}{c}{Target Behavior} & \multicolumn{2}{c}{Behavior Specific} & \multicolumn{2}{c}{Behavior-Item} \\
            \cmidrule(lr){2-3} \cmidrule(lr){4-5} \cmidrule(lr){6-7} \cmidrule(lr){8-9} \cmidrule(lr){10-11} \cmidrule(lr){12-13}
            & NDCG@5 & NDCG@10 & NDCG@5 & NDCG@10 & NDCG@5 & NDCG@10 & NDCG@5 & NDCG@10 & NDCG@5 & NDCG@10 & NDCG@5 & NDCG@10\\ 
            \midrule
            \multicolumn{13}{c}{\centering \emph{PBA Transformer Modules Ablation}}\\
            \midrule
            \emph{w/o} PR & 0.3765 & 0.3894 & 0.1977 & 0.2050 & \underline{0.1643} & \underline{0.1730} & \underline{0.3325} & \underline{0.3778} & 0.2897 & \underline{0.3188} & 0.2122 & 0.2258\\ 
            \emph{w/o} BI & \underline{0.3818} & \underline{0.3970} & 0.1946 & 0.2030 & 0.1593 & 0.1697 & 0.3138 & 0.3575 & 0.2756 & 0.3038 & 0.2029 & 0.2159\\ 
            \emph{w/o} PR \& BI & \textbf{0.3853} & \textbf{0.3999} & \underline{0.1979} & 0.2059 & 0.1633 & 0.1729 & 0.3155 & 0.3607 & 0.2743 & 0.3036 & 0.1945 & 0.2085\\
            \midrule
            \multicolumn{13}{c}{\centering \emph{Item Tokenizer Ablation}}\\
            \midrule
            HSTU & N/A & N/A & N/A & N/A & 0.1579 & 0.1616 & N/A & N/A & N/A & N/A & 0.1607 & 0.1654\\ 
            \emph{w/o} IT & 0.3565 & 0.3598 & 0.1946 & 0.1959 & 0.1585 & 0.1637 & 0.3318 & 0.3396 & 0.2875 & 0.2926 & \underline{0.2373} & \underline{0.2506}\\ 
            TIGER & 0.2049 & 0.2263 & N/A & N/A & N/A & N/A & 0.2595 & 0.2788 & N/A & N/A & N/A & N/A\\ 
            RQ-VAE ($128^4$) & 0.3690 & 0.3861 & 0.1951 & \underline{0.2074} & 0.1592 & 0.1715 & 0.2993 & 0.3265 & 0.2649 & 0.2833 & 0.1853 & 0.1947\\ 
            RQ-VAE ($256^4$) & 0.3460 & 0.3634 & 0.1665 & 0.1799 & 0.1315 & 0.1445 & 0.3323 & 0.3684 & \underline{0.2909} & 0.3156 & 0.2008 & 0.2149\\ 
            \midrule
            \multicolumn{13}{c}{\centering \emph{Ours}}\\
            \midrule
            \textbf{MBGen (SID)}& 0.3800 & 0.3954 & \textbf{0.2053} & \textbf{0.2131} & 0.1712 & \textbf{0.1803} & \textbf{0.3498} & \textbf{0.3949} & \textbf{0.3023} & \textbf{0.3317} & 0.2155 & 0.2301 \\
            \textbf{MBGen (CID)}& 0.3815 & 0.3934 & 0.2030 & 0.2076 & \textbf{0.1720} & 0.1800 & 0.3454 & 0.3913 & 0.2972 & 0.3261 & \textbf{0.2443} & \textbf{0.2730}\\
            \bottomrule
        \end{tabular}
        }
        \vspace{-0.1in}
    \end{table*}
\subsection{Ablation Study}
We analyze how each of the proposed components affects final performance. \Cref{tab:ablation_exp} shows the performance of the two full variants of our proposed MBGen model and eight models with some key designs removed. 

\subsubsection{\textbf{PBA Transformer Module Ablation}}\

(1) \emph{w/o} Position Routed Sparse FFN: The position routed sparse structure in the PBA Transformer model helps to scale up the model without an increase in inference cost. We see a degradation in performance due to the lack of model capacity, as expected.

(2) \emph{w/o} Behavior Injection: Removing the behavior injection module undermines the model's ability to capture the semantics of the behaviors at a token level. 

(3) \emph{w/o} Position Routed Sparse \& FFN Behavior Injection: The ablation of both modules could largely hurt the model's capacity and ability to capture fine-grained behavior patterns. This observation is even more pronounced in the larger-scale IJCAI dataset.

\subsubsection{\textbf{Tokenizer Ablation}}\

(1) HSTU Tokenizer: The tokenizer proposed in HSTU~\cite{zhai2024actions} tokenizes an interaction into an item token followed by a behavior token. As mentioned, we argue that this method doesn't adhere to the two-step nature of the MBSR problem. We observe a worse performance than the \emph{w/o} item tokenizer variant due to its reversed behavior and item token order. Additionally, such a tokenizer also lacks the capability of conditioning next-item prediction on known behavior types.

(2) \emph{w/o} Item Tokenizer: We remove the item tokenizer and replace the original 3 item tokens with one single item token. The model cannot model the token-level behavior-item interaction patterns without the item tokenizer. When the item tokenizer is removed, it's also harder to retrieve the next-item prediction from large amount of candidates. We do observe that this variant with the item tokenizer removed outperforms our proposed Semantic ID variant, and we hypothesize that this is caused by the complicated probability distribution when we retrieve top candidates using beam search. We detail this discussion in~\Cref{sec:BA_sampling}.

(3, 4, 5) RQ-VAE Tokenizer: We test the behavior-aware RQ-VAE tokenizer on two codebook sizes, which are both larger than the codebook size of the proposed balanced item tokenizer due to the above-mentioned code imbalance nature of the RQ-VAE tokenizer. Thus, an even smaller RQ-VAE codebook size is impractical due to the very serious collision problem. Both the code imbalance, the bigger codebook size, and more codebook layers make it hard for the model to perform next-item prediction. We perform quantitative analysis and further discuss the code distribution of the RQ-VAE tokenizer and our proposed balanced tokenizer in~\Cref{sec:code_distribution}. Additionally, we also compare the behavior-aware RQ-VAE tokenizer with the original TIGER tokenizer without behavior-aware designs, which demonstrates the role that behavior token plays in our model. In addition to the incapability of behavior-specific and behavior-item predictions, we also observe a big performance decay in target behavior prediction tasks. This shows the effectiveness of our designed behavior-aware tokenizer schema.

\begin{table}[t]
    \centering
    \caption{Quantitative analysis on the code distribution of different item tokenizers. The lower the variance and collision, the more balanced the generated codes are.}
    \label{tab:code_distribution}
    \resizebox{\linewidth}{!}{
    \begin{tabular}{ccccc}
        \toprule
        \multirow{2}{*}{Metrics} & \multicolumn{2}{c}{Retail} & \multicolumn{2}{c}{IJCAI} \\
        \cmidrule(lr){2-3} \cmidrule(lr){4-5}
            & Balanced SID & RQ-VAE SID & Balanced SID & RQ-VAE SID \\
        \cmidrule(lr){1-1} \cmidrule(lr){2-3} \cmidrule(lr){4-5}
            L1 Variance & 90,093 & 152,325 & 3,651,624 & 7,562,833 \\ 
            L2 Variance & 71 & 238 & 601 & 2,487 \\ 
            L3 Variance & 0.2368 & 0.6778 & 0.024 & 1.15 \\ 
            L4 Collisions & N/A & 51,083 & N/A & 220,263\\ 
        \bottomrule
        \end{tabular}
    }
\end{table}

\begin{figure*}[t]
    \centering
    \begin{minipage}{0.455\linewidth}
    \includegraphics[width=\linewidth]{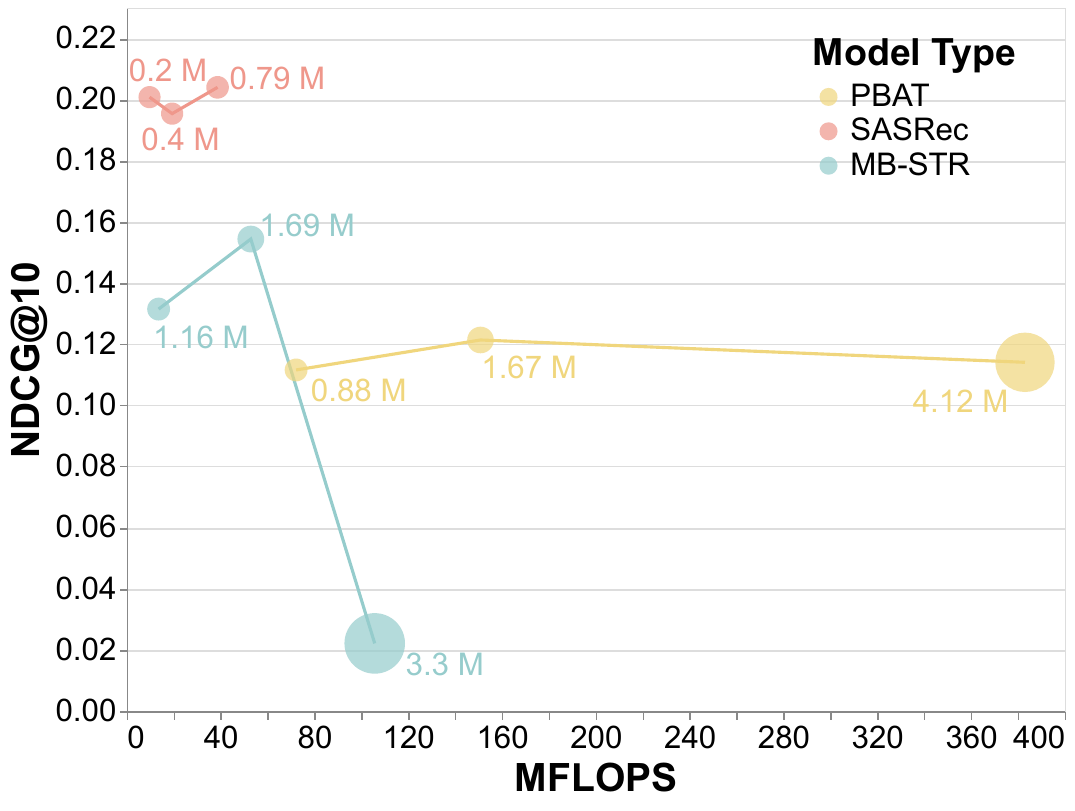}
    \caption{Baseline models scalability comparison on target behavior item prediction task on IJCAI dataset. The model computational budget is measured from one single forward call on inference.}
    \label{fig:scaling_baseline}
    \end{minipage}
    \hfill
    \begin{minipage}{0.478\linewidth}
    \includegraphics[width=\linewidth]{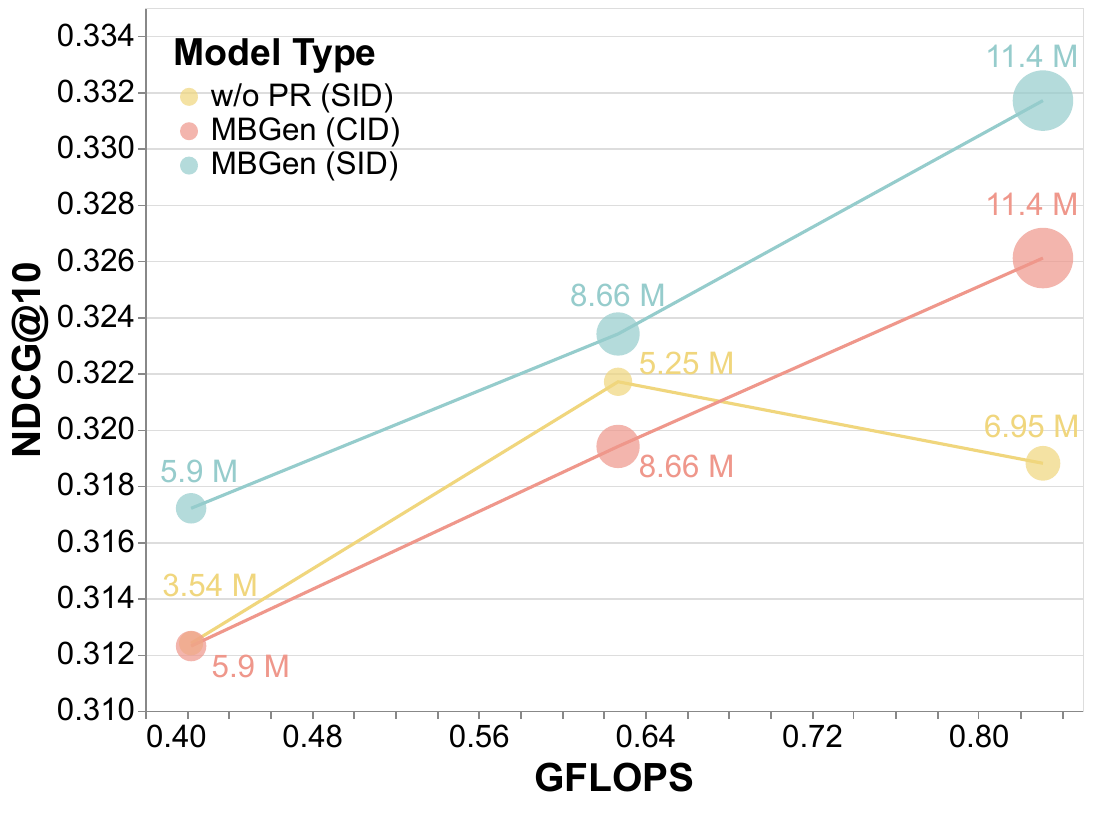}
    \caption{MBGen variants scalability comparison on behavior-specific item prediction task on IJCAI dataset. The model computational budget is measured from one single forward call on inference.}
    \label{fig:scaling_main}
    \end{minipage}
\end{figure*}

\begin{figure}[t]
    \centering
    \begin{minipage}{0.49\linewidth}
    \includegraphics[width=\linewidth]{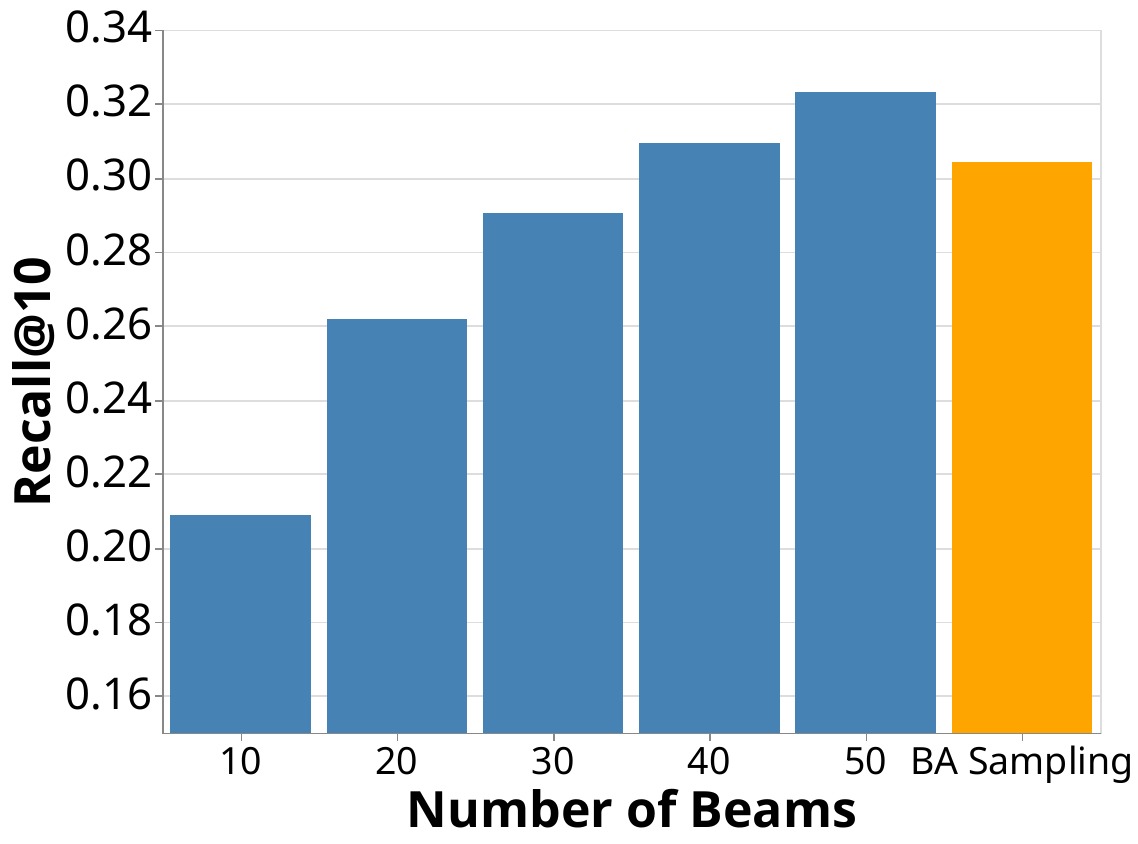}
    \end{minipage}
    \hfill
    \begin{minipage}{0.49\linewidth}
    \includegraphics[width=\linewidth]{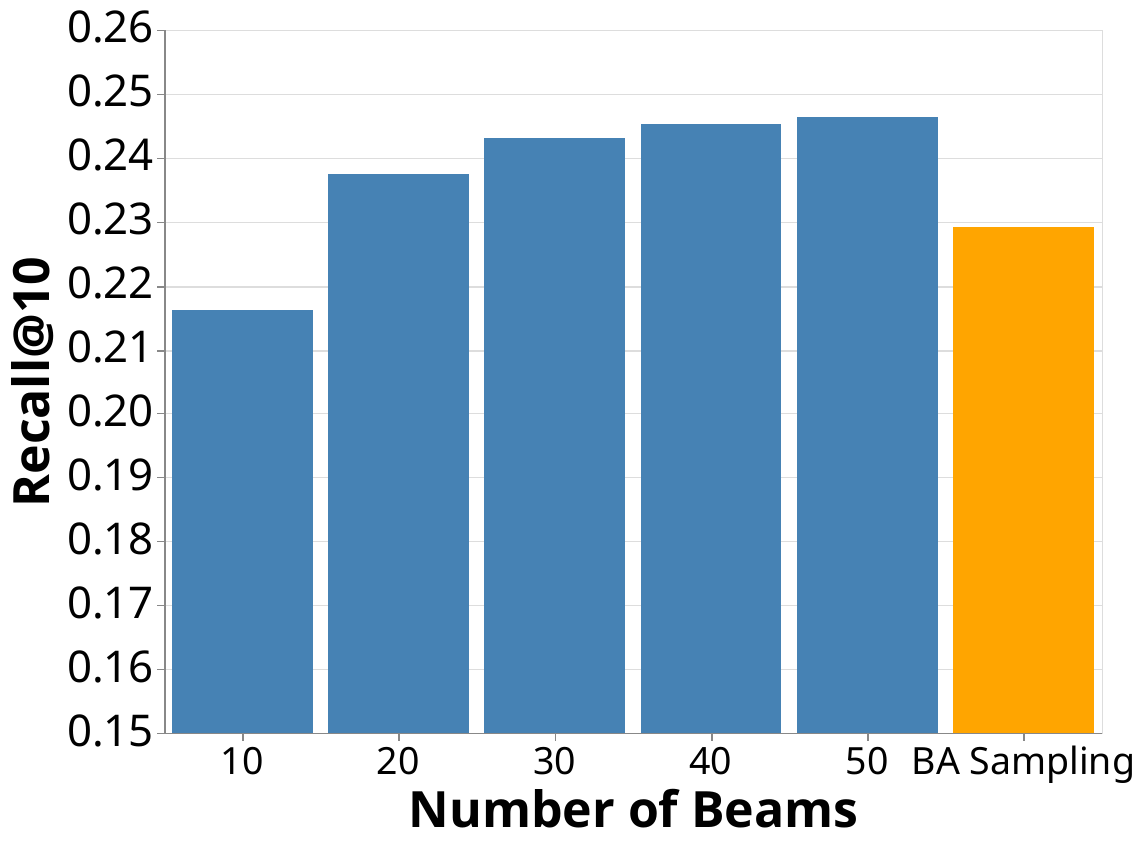}
    \end{minipage}
    \caption{Comparison of retrieval performance with respect to different numbers of beams used in beam search along with the performance of proposed behavior aware sampling method on IJCAI (left) and Retail (right) datasets.}
    \label{fig:beam_search}
\end{figure}
\subsection{Further Analysis}
    \subsubsection{\textbf{Quantitative Analysis on Code Distribution}} \label{sec:code_distribution}
    We perform a quantitative analysis on the empirical distribution, and the results are shown in~\Cref{tab:code_distribution}. We omit the analysis for the Chunked IDs variant since it theoretically grants a perfectly balanced code distribution. For a fair comparison, we fit RQ-VAE models with the same codebook size as the codebook sizes of our proposed balanced semantic IDs. For the codebook layer $m$, we count the number of items with the same first $m$ digits for each possible code combination. For instance, there are $64 \times 64 = 4096$ possible combinations on the first two digits for the retail dataset. We calculate the histogram of the distribution of the first $m$ digits. Then, we compare the variance of such empirical distribution (the smaller, the more balanced the code distribution). We observe that our proposed Balanced SID has a significantly more balanced code distribution in each codebook layer. Furthermore, we measure the number of collisions on the fourth layer in the RQ-VAE tokenizer; the result shows that more than half of the items in both datasets have some other items that share the same first three codes with them due to extremely unbalanced code assignment. In contrast, there are only three layers in our designed balanced item tokenizers.
    \subsubsection{\textbf{Behavior-Aware Sampling}} \label{sec:BA_sampling}
    As shown in~\Cref{fig:beam_search}, we observe that increasing the number of beams significantly improves the performance of our model on the behavior-item prediction task. We propose that this is caused by the inability to capture the complicated conditional probability distribution of the beam search algorithm. In both datasets, some behaviors (\emph{e.g.}, purchase and add-to-favorite) appear much less often than others (\emph{e.g.}, click). Such an unbalanced probability distribution can make it hard for the beam search to retrieve the behavior-item pairs that jointly have a high probability but with a rarer behavior type. To that end, we experiment with a behavior-aware sampling method that achieves considerable improvement in the retrieval task without the need for more beams in the beam search. We first record the probability of each behavior and retrieve the same proportion of items conditioned on each behavior. For instance, if the probability of the four behaviors is $[0.3, 0.42, 0.18, 0.1]$, we use the beam search to get the top $[3, 4, 2, 1]$ results for each given behavior. We then combine the results as our top-10 behavior-item prediction. As shown in~\Cref{fig:beam_search}, such a behavior-aware sampling method significantly outperforms the original beam search algorithm with 10 beams in both datasets.
    \subsubsection{\textbf{Scalability Analysis}}\label{sec:scaling}
    We show the relation between the model capacity and the model performance of baseline methods in~\Cref{fig:scaling_baseline}. We observe that the baseline models have a degradation of performance as we scale up the mode, especially the MB-STR and PBAT models. As in~\Cref{fig:scaling_main}, we observe that the variants of MBGen have better scalability when compared to the baseline models. Furthermore, we observe that our model with the position routed sparse FFN architecture performs better than the variant model without the sparse architecture with the same amount of computation consumption. 

\section{Conclusion}
In this work, we propose MBGen, the first generative recommendation on the multi-behavior sequential recommendation problem. Different from previous models that only model the second step (next item prediction given next behavior) of the MBSR problem, we construct a unified multi-behavior generative recommendation paradigm that's coherently trained on both steps of the MBSR problem. We further design a balanced behavior-aware item tokenizer that allows our model to learn from token-level fine-grained interaction patterns. Exploiting the heterogeneous nature of our constructed behavior-item sequence, we designed a position-routed sparse architecture to efficiently scale up the model. Extensive experiments are conducted with public datasets to show that our model significantly outperforms all baseline models in multiple MBSR tasks.

\bibliographystyle{ACM-Reference-Format}
\bibliography{reference}

\appendix

\end{document}